\documentclass[journal,twocolumn]{IEEEtran}
\usepackage{graphicx}
\usepackage{epstopdf}
\usepackage{diagbox}
\usepackage{amssymb}
\usepackage{amsfonts}
\usepackage{amsmath}
\usepackage{algorithm}
\usepackage{algorithmic}
\usepackage{subeqnarray}
\usepackage{cases}

\usepackage{graphicx,subfigure,amsmath,amssymb,cite}
\usepackage{float}
\ifCLASSINFOpdf
\else
\fi
\begin{document}

\title{High-performance Passive Eigen-model-based Detectors of Single Emitter Using Massive MIMO Receivers}

\author{Qijuan Jie, Xichao Zhan,~Feng Shu,~Yaohui Ding, Baihua Shi, Yifan Li,~and Jiangzhou Wang,\emph{ Fellow, IEEE}

\thanks{Qijuan Jie,~Xichao~Zhan, Feng Shu,~and~Yaohui~Ding are with the School of Information and Communication Engineering, Hainan Unversity, Haikou, 570228, China(e-mail: shufeng0101@163.com). }
\thanks{Baihua Shi~,Yifan Li, and Feng Shu are with School of Electronic and Optical Engineering, Nanjing University of Science and Technology, Nanjing, 210094, China.}
\thanks{Jiangzhou Wang is with the School of Engineering and Digital Arts, University of Kent, Canterbury CT2 7NT, U.K. Email: j.z.wang@kent.ac.uk.}


}

\maketitle

\begin{abstract}
 For a passive direction of arrival (DoA) measurement system using massive multiple input multiple output (MIMO), it is mandatory to infer whether the emitter exists or not before performing DOA estimation operation. Inspired by the detection idea from radio detection and ranging (radar), three high-performance detectors are proposed to infer the existence of single passive emitter from the eigen-space of sample covariance matrix of receive signal vector. The test statistic (TS) of the first method is defined as the ratio of maximum eigen-value (Max-EV) to minimum eigen-value (R-MaxEV-MinEV) while that of the second one is defined as the ratio of Max-EV to noise variance (R-MaxEV-NV). The TS of the third method is the mean of maximum eigen-value (EV) and minimum EV(M-MaxEV-MinEV). Their closed-form expressions are presented and the corresponding detection performance is given. Simulation results show that the proposed M-MaxEV-MinEV and R-MaxEV-NV methods can approximately achieve the same detection performance that is better than the traditional generalized likelihood ratio test method with false alarm probability being less than 0.3.



\end{abstract}
\section{Introduction}
Antenna direction-finding (DF) technology has been widely used in diverse engineering fields, including radar, sonar, navigation, tracking of various objects and beamforming track and alignment in fifth generation/sixth generation (5G/6G) utilizing massive or ultra-massive multiple input multiple output (MIMO) \cite{Tuncer2009Classical}. In the future, several emerging engineering applications will also present opportunities for applying DF, including the internet of things (IoT), direction modulation (DM) \cite{2019Multi}-\cite{2019WFRFT}, unmanned aerial vehicle communications, intelligent transportation, and wireless sensor networks etc. In recent years, with the emergence of massive antenna array structures, as a traditional field, the direction of arrival (DOA) estimation begin to be a new life.  In \cite{shuDOA}, three low-complexity and high-resolution methods were proposed for hybrid massive MIMO systems: two parametric methods and a root-MUSIC-based method. At the same time, the Cramer-Rao lower bound (CRLB) for hybrid massive MIMO was also presented as a performance metric. \cite{2019One} has made an investigation of the DOA estimation problem in the presence of 1-bit analog-to-digital converters (ADCs) and proved that the multiple signal classification (MUSIC) method can be directly extended to this case without additional preprocessing and analyzed the system performance degradation when the bit length is 1. Authors in \cite{2020DOA} proposed  a root-propagator method based on a uniform linear antenna array for DOA estimation and computed the range of the near-field source without prior knowledge of the source number. The results showed that compared with other existing similar methods, the proposed method provided superior performance in near-field source positioning. A DOA estimation and channel estimation scheme based on deep learning (DL) was proposed in \cite{2018Deep}, and the proposed scheme achieved better performance in both DOA estimation and channel estimation. In \cite{2019Low}, the authors presented an improved low-complexity DL method to a hybrid massive MIMO system with uniform circular array. This method is similar to the traditional maximum likelihood estimation (ML) method with even better performance and lower complexity.Large-scale means a large number of receiving antennas, ADCs and radio frequency (RF) links etc, which will cause a significant increase in circuit costs. In order to overcome this challenge problem, in \cite{2021On}-\cite{2017Hybrid}, the DOA estimation performance of a massive MIMO receiving array with a hybrid ADC structure was studied, which can achieve a good balance between the root mean square error, circuit cost and energy efficiency. Furthermore, in\cite{2013Nonlinear} authors proposed a ML method in a specific application. Stoica and Sharman combined ML and MUSIC to construct five methods \cite{1990Maximum}, which achieved good performance.

In \cite{2007An}, authors integrated general problems of signal detection, and focused on the likelihood ratio test(LRT) rule. Authors in \cite{2012Target} proposed to detect target with MIMO radar and explored many target detection algorithms, by eigenvalue decomposition. The independent observation components were extracted from $M$ orthogonal signals and weighted together. Therefore target detection method is preferred for more practical applications.

To the best of our knowledge, passive DOA measurement has not been studied so far. The main research efforts focused on DOA measurement methods, CRLB, and array calibration etc. It is very necessary to know whether the emitter exists or not before making a DOA estimation.  In this paper, we have proposed and investigated the passive detection of single emitter. The main contributions in this paper are summarized as follows:

\begin{enumerate}
\item To achieve a high detection performance of weak emitter and trigger the next step: DOA measurements, an EVD-based detection method, called R-MaxEV-MinEV, is proposed. Here, the sampling covariance of receive signal vector is computed, and its EVD is performed to extract all its eigenvalues. The test statistic is defined ,as the ratio of the largest eigen-value to smallest eigen-value. Simulations results show that the proposed R-MaxEV-MinEV is slightly better than traditional generalized likelihood ratio test (GRLT) in terms of detection probability.

\item To enhance the detection performance, two novel detection methods are proposed, called M-MaxEV-MinEV and R-MaxEV-NV. The test statistic of M-MaxEV-MinEV is defined as the arithmetic mean (AM) of the maximum value and minimum one while that of R-MaxEV-NV is defined as the ratio of the maximum eigen-value and the estimated noise variance. The corresponding noise variance is estimated using the average value of all eigenvalues excluding the largest one. Simulation results show that the proposed M-MaxEV-MinEV and R-MaxEV-NV are much better than existing GLRT and R-MaxEV-MinEV in accordance with ROC and detection probability. In particular, it is noted that the improved detection performance gains of the proposed methods over GLRT are  achieved at the expense of an additional computational complexity of EVD.
\end{enumerate}
The remainder of this paper is organized as follows. Section II describes the system model of passive detection of weak emitter using massive MIMO. In Section III, three detectors are proposed, and their performance and computational complexities are also analyzed. We present our simulation results in Section IV. Finally, we draw conclusions in Section V.

\emph{Notations:} Throughout the paper, $\mathbf{x}$ and $\mathbf{X}$ in bold typeface are used to represent vectors and matrices, respectively, while scalars are presented in normal typeface, such as $x$. Signs $(\cdot)^H$ and $|\cdot|$ represent conjugate transpose and modulus, respectively. $\mathbf{I}_N$ denotes the $N\times N$ identity matrix. Furthermore, $\mathbb{E}[\cdot]$ represents the expectation operator, and $\mathbf{x}\sim \mathcal{CN}(\mathbf{m},\mathbf{R})$ denotes a circularly symmetric complex Gaussian stochastic vector with mean vector $\mathbf{m}$ and covariance matrix $\mathbf{R}$. $\mathbf{diag}(\mathbf{X})$ denotes a diagonal matrix by keeping only the diagonal elements of matrix $\mathbf{X}$. $\mathbf{Tr}(\cdot)$ denotes matrix trace. $\hat{x}$ represents the estimated value of $x$.

\section{system model}
\begin{figure}[h]
\centering
\includegraphics[width=0.51\textwidth]{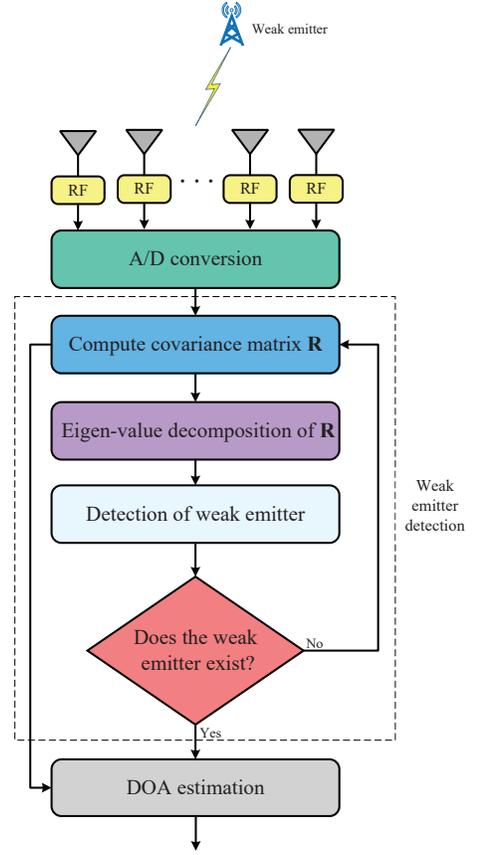}\\
\caption{Massive DOA measurements with the ability of detecting a weak emitter.}\label{system-model.eps}
\end{figure}
In Fig.\ref{system-model.eps}, a block diagram of DOA measurements with the ability of detecting a weak emitter is sketched. Here, the narrow-band signal $s(t)e^{j2\pi f_ct}$ of a far-field weak emitter impinges on the array, where $s(t)$ represents the baseband signal and $f_{c}$ represents the carrier frequency. We consider a uniform linear array (ULA) and the ULA has $N$ antenna elements. Moreover, different received antennal elements will capture the signal at different time, which are expressed as:
\begin{align}\label{y(t)}
\widetilde{\textbf y}(t)=&[s(t-\tau_{1})e^{j2\pi f_{c}(t-\tau_{1})}, \nonumber\\
&s(t-\tau_{2})e^{j2\pi f_{c}(t-\tau_{2})}, \cdots, s(t-\tau_{N})e^{j2\pi f_{c}(t-\tau_{N})}]^{T}
\end{align}
where $\tau_{m}$ stands for the delay from the emitter to the $m$th antenna of the receiving array, for a linear array, $\tau_{m}=\tau_{0}-(d_{m}/c){sin}\theta$. Here,  $\tau_{0}$ represents the propagation delay from the emitter to a reference point on the array, $d_m,~ m=1,2,\cdots,N$, denotes the distance from the $m$th antenna to the reference point, $c$ represents the speed of light, and $\theta_{0}$ is the direction of the emitter relative to the line perpendicular to the array. It is generally assumed that $\tau_{0}=0$, $\tau_{m}=-(d_{m}/c){sin}\theta$ at this time. Since there is $s(t-\tau_{m})\approx s(t)$ for a narrowband signal, the passband signal in (\ref{y(t)}) is converted into a baseband signal
\begin{align}
\textbf{y}(t)=s(t)[e^{j\omega_{c}\tau_{1}}, e^{j\omega_{c}\tau_{2}}, \cdots, e^{j\omega_{c}\tau_{N}}]^{T}
\end{align}
where $\omega_{c}=2\pi f_{c}$. Let us define the array manifold
\begin{align}
{a}(\theta_{0})=[e^{j2\pi d_{1}{sin}\theta/\lambda}, e^{j2\pi d_{2}{sin} \theta/\lambda}, \cdots, e^{j2\pi d_{N}{sin}\theta/\lambda}]^{T},
\end{align}
then the model of unit power signal received by the antenna array is
\begin{align}
\textbf{y}(t)=\sqrt{\operatorname{SNR}}\textbf a(\theta_{0}){s}(t)+\textbf{v}(t),
\end{align}
where $\textbf{v}(t)$ is the multivariate complex Gaussian noise vector, $SNR$ is the signal-to-noise ratio, and $a(\theta _{0})$ is the signal transmission direction. The output of the antenna array is generally expressed in the discrete form of sampling points
\begin{align}
\textbf{y}[k]=\sqrt{\operatorname{SNR}}\textbf a(\theta_{0}){s}[k]+\textbf{v}[k]
\end{align}
where $k$ represents the $k$th sampling point. Thus, through RF chains, the signal is down-converted and the received signal vector can be rewritten as
\begin{equation}\label{y}
\mathbf{y}(t)=\mathbf{a}(\theta_{0})s(t)+\mathbf{v}(t),
\end{equation}
where $\mathbf{v}(t)\sim\mathcal{CN}(0,\mathbf{I}_N)$ is the additive white Gaussian noise (AWGN). Collecting all $N$ sampling vectors forms the following receive signal matrix. The signal s(t) is independent of v(t).

It is assumed that the signal $\textbf s(l)$ is the baseband signal, $E[|\textbf s(l)|^{2}]=P_{S}$, its mean is a vector of all-zero with a diagonal nonsingular semipositive definite covariance matrix $\textbf R_{s}$, $\textbf R_{\textbf s}=E\{ \textbf s(l)\textbf s^{H}(l)\}\geq 0$. The mean of noise vector $\textbf v(l)$ is a vector of all-zero and is both spatially and temporally white $E[\textbf v(l)\textbf v^{H}(l)]=\sigma_{\textbf v}^{2}\textbf I_{N}$. In the presence of single emitter, the covariance matrix of $\mathbf{R}_{\mathbf{y}}$ is given by
\begin{align}
\mathbf{R}_{\mathbf{y}}&=\mathbf{diag}(\sigma_s^2\mathbf{a}(\theta_{0}) \mathbf{a}^H(\theta_{0})+\sigma_{\textbf v}^{2}\mathbf{I}_{N})\nonumber\\
&=(\sigma_s^2+\sigma_{\textbf v}^{2})\mathbf{I}_{N}.
\end{align}

The detection problem of weak emitter is viewed as the following binary hypothesis problem
\begin{equation}
\left\{
             \begin{array}{lr}
             H_{0}: &~~~~  y(t)=v(t),~~~~~\\
             H_{1}: &  y(t)=a(\theta)s(t)+v(t),
             \end{array}
\right.
\end{equation}
which yields the GLRT as follows
\begin{equation}
T_{GLRT}=\frac{p(y|H_{1})}{p(y|H_{0})}\underset{{H}_{0}}{\overset{{H}_{1}}\gtrless} \gamma_{GLRT},
\end{equation}
which further gives
\begin{equation}
T_{1}=\frac {\frac{1}{N}\sum_{n=1}^{N}\lambda_{n}}{(\prod_{i=1}^{N}\lambda_{i})^{\frac{1}{N}}}=\frac{\frac {1}{N}\mathbf{Tr}(\mathbf{R}_{\mathbf{y}})}{(\mathbf{det}(\mathbf{R}_{\mathbf{y}})^{\frac{1}{ N}}}=\underset{{H}_{0}}{\overset{{H}_{1}}\gtrless}\gamma_{1}.
\end{equation}

\section{Proposed three passive detectors}
In this section, the EVD of sampling covariance matrix is performed and its eigen-value joint density function is given directly. Thus, three high-performance detectors are proposed to detect the weak emitter.
\subsection{EVD and joint density function of eigenvalues of sampling covariance matrix}
%
%

In practice, although the covariance matrix $\mathbf{R}_{\mathbf{y}}$ can not be obtained, its estimated value is given by
\begin{equation}\label{R_y}
\hat{\mathbf{R}}_{\mathbf{y}}=\frac{1}{L}\sum\limits_{n=1}^{L}\mathbf{y}(n)\mathbf{y}^{H}(n).
\end{equation}
where $L$ is the number of snapshots. Letting
\begin{equation}\label{A}
\mathbf{A}=L\hat{\mathbf{R}}_{\mathbf{y}}=\sum\limits_{n=1}^{L}\mathbf{y}(n)\mathbf{y}^{H}(n),
\end{equation}
we have
\begin{equation}\label{p(y)}
P(\mathbf{y})=\frac{1}{\pi^{N}|\mathbf{R}_{\mathbf{y}}|}e^{-{\mathbf{y}^{H}}\mathbf{R}_{\mathbf{y}}^{-1}\mathbf{y} }
\end{equation}

According to the Wishart matrix theorem in \cite{Goodman1963Statistical}, $\mathbf{A}$ is a random Wishart matrix, with the Wishart distribution \begin{equation}\label{p(A)}
P_{W}(\mathbf{A})=[|\mathbf{A}|^{L-N}/\mathbf{I}(\mathbf{R}_{\mathbf{y}})]exp[-\mathbf{Tr}(\mathbf{R}_{\mathbf{y}}^{-1}\mathbf{A})]
\end{equation}
where
\begin{equation}\label{I(Ry)}
\mathbf{I}(\mathbf{R}_{\mathbf{y}})=\pi^{\frac{1}{2}N(N-1)}[\Gamma(L)\cdot\cdot\cdot\Gamma(L-N+1)|\mathbf{R}_{\mathbf{y}}|^{L}
\end{equation}
$\hat{\mathbf{R}}_{\mathbf{y}}$ follows the Wishart distribution with $L$ degrees of freedom and covariance matrix $\sigma_{v}^{2}\mathbf{I}_{N}/L $
\begin{equation}
\hat{\mathbf{R}}_{\mathbf{y}}\sim W_{N}(L,\sigma_s^2\mathbf{a}(\theta_{0}) \mathbf{a}^H(\theta_{0})/L+\sigma_{v}^{2}\mathbf{I}_{N}/L)
\end{equation}

%
According to Theorem (1.6) in \cite{Goodman1963Statistical}, we have the joint density function of  $\hat{\mathbf{R}}_{\mathbf{y}}$
\begin{equation}\label{p(Ry)}
P_{W}(\hat{\mathbf{R}}_{\mathbf{y}})=[|\mathbf{A}|^{L-N}/\mathbf{I}(\mathbf{R}_{\mathbf{y}})]exp[-\mathbf{Tr}(\mathbf{R}_{\mathbf{y}}^{-1}\hat{\mathbf{R}}_{\mathbf{y}})]
\end{equation}

Let us define $\lambda_{1}, \lambda_{2},...,\lambda_{N}$ as $N$ non-negative eigenvalues of $\hat{\mathbf{R}}_{\mathbf{y}}$. Since $\hat{\mathbf{R}}_{\mathbf{y}}$ is hermitian and positive semidefinite, we have its EVD as
\begin{align}\label{R_y-EVD}
\hat{\mathbf{R}}_{\mathbf{y}}=\mathbf{U}\boldsymbol{\Sigma}\mathbf{U}^ { H }, ~\boldsymbol{\Sigma}=\text{diag}\{\lambda_n\}^N_{n=1},~\lambda_n\in \mathbb{R}_{+}\cup\{0\}
\end{align}

The joint PDF of all eigen-values are as follows
\begin{align}\label{p_lambda1}
&p_{N,L}^{Wishart}(\lambda_{1}, \lambda_{2}, \ldots, \lambda_{N}) \nonumber\\
&=c_{N,L}^{-1}\prod_{1\leq i<j \leq N}|\lambda_{i}-\lambda_{j}|\prod_{i=1}^ {N}\lambda_{i}^{\alpha/2}e^{-\lambda i/2}
\end{align}
where $c_{N,L}$ is a constant depending only on $N$ and $L$. $\alpha=L-N-1$, $\lambda_{1}\geq \lambda_{2}\cdots\geq\lambda_{N} $ is the eigenvalue sequence of $\hat{\mathbf{R}}_{\mathbf{y}}$. Let $\lambda_{max}$ and $\lambda_{min}$ denote the maximum and minimum eigen-values of $\hat{\mathbf{R}}_{\mathbf{y}}$
%
%


According to the M-P Law, as $L\rightarrow\infty$ with $N/L=c, c\in(0,1)$, we have the following limit results
\begin{equation}
\begin{aligned}
\lambda_{min}\rightarrow a=\frac{1}{L}(\sqrt{L}-\sqrt{N})^{2}\\
\lambda_{max}\rightarrow b=\frac{1}{L}(\sqrt{L}+\sqrt{N})^{2}
\end{aligned}
\end{equation}

To obtain the exact probability density function (pdf) of the smallest eigenvalue of the Wishart matrix, we rewrite (\ref{p_lambda1}) as
\begin{align}\label{p_lambda2}
&\hat{p}_{N,L}^{Wishart}(\lambda_{1}, \lambda_{2}, \ldots, \lambda_{N}) \nonumber\\
&=d_{N,L}\prod_{1\leq i<j \leq N}|\lambda_{i}-\lambda_{j}|\prod_{i=1}^ {N}\lambda_{i}^{\alpha/2}e^{-\lambda i/2}d\lambda_{1}\ldots d\lambda_{N}
\end{align}
where $d_{N,L}$ is a constant depending only on $N$ and $L$.

Using symmetry, integrating (\ref{p_lambda2}) over the remaining eigenvalues generates the marginal pdf for the smallest eigenvalue $\lambda_{min}$
\begin{small}
\begin{align}\label{f_lambdamin}
&f_{\lambda_{min}}(\lambda)=\frac{d_{N,L}}{(N-1)!}\lambda^{L-N-1/2}e^{-\lambda/2}\nonumber\\
&\int_{R_{\lambda}}e^{-\sum\limits_{i=1}^{N-1}\frac{\lambda_{i}}{2}}\prod_{1\leq i<j\leq L-1}|\lambda_{i}-\lambda_{j}| \prod_{i=1}^{L-1}(\lambda_{i}-\lambda)\lambda_{i}^{\frac{L-N-1}{2}}d\lambda_{i}
\end{align}
\end{small}
where $R_{\lambda}=\{(\lambda_{1}, \ldots, \lambda_{N-1}): \lambda_{i}>\lambda\}$.

Now, performing the change of variables $x_{i}=\lambda_{i}-\lambda$, we obtain
\begin{align}\label{p_lambda3}
&f_{\lambda_{min}}(\lambda)=
\frac{d_{N,L}}{(N-1)!}\lambda^{\frac{L-N-1}{2}}e^{-\lambda N/2}\nonumber\\
&\times\int_{R_{+}^{N-1}} \prod_{i=1}^{N-1}(x_{i}+\lambda)^{\frac{L-N-1}{2}}\Delta d \Omega
\end{align}
where $\Delta=\prod_{1\leq i<j\leq L-1}|x_{i}-x_{j}|$, $d\mu(x)=xe^{-x/2}dx$,
$d\Omega=d\mu(x_{1})\ldots d\mu(x_{L-1})$, and $R_{+}^{N-1}=\{(x_{1}, \ldots, x_{N-1}): x_{i} \geq 0\}$.
Similarly, the marginal PDF of the largest eigenvalue is directly given by
\begin{align}
&f_{\lambda_{\max}}(\lambda)\leq a_{N,L}\lambda^{\frac{1}{2}(L+N-3)}e^{-\lambda/2}\nonumber\\
&=\frac{\pi^{1 /2} 2^{\frac{1-L-N}{2}}}{\Gamma(L/2) \Gamma(N/2)} \lambda^{\frac{1}{2}(L+N-3)} e^{-\lambda/2}
\end{align}
\subsection{Proposed R-MaxEV-MinEV}
The proposed ratio detector, R-MaxEV-MinEV,is defined as  the ratio of the largest eigen-value to smallest eigen-value, and the corresponding test statistic is
\begin{equation}
T_{2}=\frac{\lambda_{max}}{\lambda_{min}}\underset{{H}_{0}}{\overset{{H}_{1}}\gtrless} \gamma_{2}
\end{equation}
It is noted that both the maximum and minimum eigenvalues of $\hat{\mathbf{R}}_{\mathbf{y}}$ obey the first-order Tracy-Widom distribution, and the description of the distribution based on the minimum eigenvalue is more accurate than that of the maximum eigenvalue, especially in the case of lower dimensions. Assuming $\operatorname{lim}_{L\rightarrow\infty}N/L=c\in(0,1)$, the maximum eigenvalue of  matrix $A(L)$ is denoted by $\lambda_{max}(A(N))$. Then $[\lambda_{ max}(A(N))-\mu]/v$ converges to the Tracy-Widom distribution of order 1, $F_{1}(t)$ in probability. To achieve a CFAR, the distribution of the linear transformation of the maximum eigenvalue $\lambda_{max}$ is adopted to deduce the relationship between threshold $\gamma_{2}$ and the false alarm probability (FAP) $P_{F}$. The associated deriving process is as follows:
\begin{equation}
\begin{aligned}
P_{F} &=P(\frac{\lambda_{\max }}{\hat{\sigma}_{v}^{2}}>\gamma_{3} \mid H_{0})=P(\lambda_{\max }>\hat{\sigma}_{v}^{2} \gamma_{3} \mid H_{0}) \\
&=P(\frac{\lambda_{\max }-\mu}{v}>\frac{\hat{\sigma}_{v}^{2} \gamma_{3}-\mu}{v})=1-F_{1}(\frac{\hat{\sigma}_{v}^{2} \gamma_{3}-\mu}{v})
\end{aligned}
\end{equation}
namely,
\begin{equation}
\gamma_{2} =\frac{\lambda_{max}} {F_{1}^{-1}(1-P_{F})\mu+v}
\end{equation}
\subsection{Proposed R-MaxEV-NV and M-MaxEV-MinEV}
The proposed second ratio detector, R-MaxEV-NV, is defined to be the ratio of the maximum eigenvalues to noise variance. Observing (\ref{R_y-EVD}), there exists only one emitter. It is evident that the remaining eigenvalues excluding the maximum eigen-value are noise variance, and a good estimator of noise variance is
\begin{align}\label{R_y-EVD}
\hat\sigma_v^2=\frac{1}{N-1}\sum_{n=2}^N\lambda_n
\end{align}
which yields the following ratio test
\begin{equation}
T_{3}=\frac {\lambda_{max}} {\hat\sigma_{v}^{2}} \underset{{H}_{0}}{\overset{{H}_{1}}\gtrless} \gamma_{3}
\end{equation}
which gives us the corresponding FAP
\begin{equation}
\begin{aligned}
P_{F} &=P(\frac{\lambda_{max}} {\hat\sigma_v^2}> \gamma_{3}|H_{0})=P({\lambda_{ max }}> {\hat\sigma_v^2}\gamma_{3}|H_{0})\\
&=P(\frac {\lambda_{max}-\operatorname\mu} {v}>\frac{{\hat\sigma_v^2}\gamma_{3}-\mu } {v})=1-F_{1} (\frac{{\hat\sigma_v^2}\gamma_{3}-\mu} {v})
\end{aligned}
\end{equation}
namely,
\begin{align}
\gamma_{3} &=\frac{F_{1}^{-1}(1-P_{F})v+\mu} {\hat\sigma_v^2}=\frac{(N-1)F_{1}^{-1}(1-P_{F})v+\mu} {\mathbf{Tr}(\hat{\mathbf{R}}_{\mathbf{y}})-\lambda_{max}}
\end{align}
The  proposed mean detector, M-MaxEV-MinEV, is simply given by
\begin{equation}
T_{4}=\frac {\lambda_{max}+\lambda_{min}}{2} \underset{{H}_{0}}{\overset{{H}_{1}}\gtrless} \gamma_{4}
\end{equation}
which gives us the corresponding FAP
\begin{equation}
\begin{aligned}
P_{F} &=P(\frac{\lambda_{max}+\lambda_{min}} {2}>\gamma_{4}|H_{0})=P(\lambda_{min}> 2\gamma_{4}-\lambda_{max})\\
&=1-F_{1} (\frac{2\gamma_{4}-\lambda_{max}-\mu}{v})
\end{aligned}
\end{equation}
which yields the threshold
\begin{equation}
\gamma_{4}=\frac{F_{1}^{-1}(1-P_{F})v+\mu+\lambda_{max}}{2}.
\end{equation}
\subsection{Complexity analysis}
Now, the computational complexities of the proposed three detectors are analyzed with the GART as a complexity benchmark.  Compared with GLRT,  the proposed three detectors achieve  a better detection performance as shown in the next section. However, they require an incremental computation about $O(N^{3}+N^2L)$ float-point operations (FLOPs), which includes computing covariance matrix and its EVD. In particular, as $N$ tends to large-scale, the complexity increment is huge.
\section{Simulation results and analysis}
In this section, we provide the practical detection performance simulation to evaluate detection performance of the three proposed detection methods. Without loss of generality, system parameters in our simulation are assumed as follows: the number of antennas $N=64$, direction-of-arrival (DOA) $\theta=\pi/6$, antennas distance $d=\lambda/2$, SNR=-18dB, the number of samples $K=200$.
\par
\begin{figure}[h]
\centering
\includegraphics[width=0.51\textwidth]{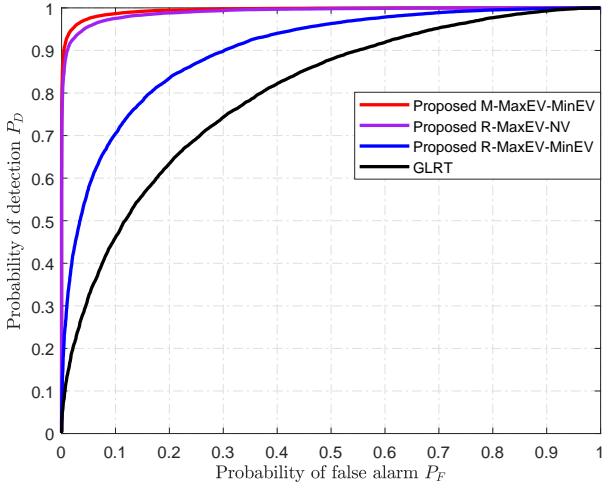}\\
\caption{Curves of ROCs of the proposed detection methods with SNR=-18dB, $N=64$, and $K=200$.}\label{roc.eps}
\end{figure}

 Fig.~\ref{roc.eps} plots the curves of receiver operator characteristic  (ROC) for the three proposed detection methods with conventional GLRT method as a performance benchmark. Compared with the existing GLRT, it can be seen from Fig.~\ref{roc.eps} that the detection performances of the proposed M-MaxEV-MinEV and R-MaxEV-NV are much better than existing GLRT for a fixed FAP. Additionally, it is noted that the proposed R-MaxEV-MinEV is slightly better than GRLT in terms of detection probability given a fixed FAP.

\par
\begin{figure}[h]
\centering
\includegraphics[width=0.51\textwidth]{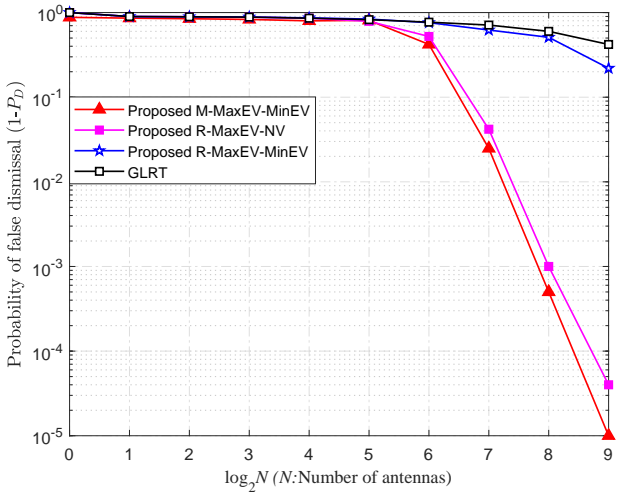}\\
\caption{False dismissal probability $P_{miss}$ under different number of antennas
  $N$ with SNR=-20dB and $P_{FA}=0.1$.}\label{N.eps}
\end{figure}

 In order to assess the tendency of the detection performance versus the number of antennas of the three proposed detection methods,  Fig.~\ref{N.eps} demonstrates  the curves of false dismissal probability (FDP) $P_{miss}$ versus  the number $N$ of antennas with $SNR$=-20dB and $P_{FA}=0.1$. It can be seen from the Figure that as $N$ increases from 2 to 512, it is very obvious that  the four methods have an increasing order on FDP performance: GRLT, R-MaxEV-MinEV,  R-MaxEV-NV,  and M-MaxEV-MinEV, where FDP is equal to one minus detection probability. Evidently, as the number of antennas goes to large-scale, the FDPs of R-MaxEV-NV  and M-MaxEV-MinE are improved dramatically.


\section{Conclusions}
In this paper, to achieve a high detection performance of weak emitter and serve the subsequent step DOA measurements, three weak emitter detectors based on eigen-model have been proposed as follows: R-MaxEV-MinEV, M-MaxEV-MinEV, and R-MaxEV-NV. The corresponding FAPs have been derived to decide the thresholds. The proposed detectors M-MaxEV-MinEV and R-MaxEV-NV performs much better than GLRT in terms of detection proability while the proposed R-MaxEV-MinEV is slightly better than GLRT. Here, the role of EVD can be viewed as a feature extraction operation similar to deep learning. In other words, the test statistic is defined on the eigen-subspaces. This leads to an improved detection performance. The performance gains over GLRT are achieved at the expense of an additional computational complexity of computing covariance matrix and its EVD. The proposed methods will be applied to the future 5G-advanced and even 6G to enable the DOA measurements and localization using mmWave and Terahertz communications in order to achieve integrated sensing and communication (ISAC).


\ifCLASSOPTIONcaptionsoff
  \newpage
\fi

\bibliographystyle{IEEEtran}
\bibliography{reference}

\begin{thebibliography}{10}
\providecommand{\url}[1]{#1}
\csname url@samestyle\endcsname
\providecommand{\newblock}{\relax}
\providecommand{\bibinfo}[2]{#2}
\providecommand{\BIBentrySTDinterwordspacing}{\spaceskip=0pt\relax}
\providecommand{\BIBentryALTinterwordstretchfactor}{4}
\providecommand{\BIBentryALTinterwordspacing}{\spaceskip=\fontdimen2\font plus
\BIBentryALTinterwordstretchfactor\fontdimen3\font minus
  \fontdimen4\font\relax}
\providecommand{\BIBforeignlanguage}[2]{{%
\expandafter\ifx\csname l@#1\endcsname\relax
\typeout{** WARNING: IEEEtran.bst: No hyphenation pattern has been}%
\typeout{** loaded for the language `#1'. Using the pattern for}%
\typeout{** the default language instead.}%
\else
\language=\csname l@#1\endcsname
\fi
#2}}
\providecommand{\BIBdecl}{\relax}
\BIBdecl

\bibitem{Tuncer2009Classical}
T.~E. {Tuncer} and B.~{Friedlander}, \emph{Classical and {Modern
  Direction}-of-{Arrival Estimation}}.\hskip 1em plus 0.5em minus 0.4em\relax
  Academic Press, 2009.

\bibitem{2019Multi}
B.~Qiu, M.~Tao, L.~Wang, J.~Xie, and Y.~Wang, ``Multi-{Beam Directional
  Modulation Synthesis Scheme Based} on {Frequency Diverse Array},'' \emph{IEEE
  Trans. Inf. Forensics Security}, vol.~14, no.~10, pp. 2593--2606, 2019.

\bibitem{2019WFRFT}
Q.~Cheng, V.~Fusco, J.~Zhu, S.~Wang, and F.~Wang, ``Wfrft-{Aided
  Power-Efficient Multi-Beam Directional Modulation Schemes Based} on
  {Frequency Diverse Array},'' \emph{IEEE Trans. Wireless Commun}, vol.~18,
  no.~11, pp. 5211--5226, 2019.

\bibitem{shuDOA}
S.~Feng, Q.~Yaolu, L.~Tingting, Z.~Yijin, L.~Jun, and H.~Zhu, ``Low-complexity
  and high-resolution {DOA} estimation for hybrid analog and digital massive
  {MIMO} receive array,'' \emph{IEEE Trans. Commun.}, vol.~58, no.~6, pp.
  2487--2501, 2018.

\bibitem{2019One}
X.~Huang and B.~Liao, ``One-{Bit MUSIC},'' \emph{IEEE Signal Process. Lett},
  vol.~26, no.~7, pp. 961--965, 2019.

\bibitem{2020DOA}
J.~Li, Y.~Wang, and Z.~Ren, ``{DOA} and {Range Estimation} using a {Uniform
  Linear Antenna Array} without a {Priori Knowledge} of the {Source Number},''
  \emph{IEEE Trans. Antennas Propag}, vol.~69, no. 2929-2939, 2021.

\bibitem{2018Deep}
H.~Huang, J.~Yang, H.~Huang, Y.~Song, and G.~Gui, ``Deep {Learning} for
  {Super-Resolution Channel Estimation} and {DOA Estimation Based Massive MIMO
  System},'' \emph{IEEE Trans. Veh. Technol}, vol.~67, no.~9, pp. 8549--8560,
  2018.

\bibitem{2019Low}
D.~Hu, Y.~Zhang, L.~He, and J.~Wu, ``Low-{Complexity Deep-Learning-Based DOA
  Estimation} for {Hybrid Massive MIMO Systems With Uniform Circular Arrays},''
  \emph{IEEE Wireless Communications Letters}, vol.~9, no.~1, pp. 83--86, 2020.

\bibitem{2021On}
\BIBentryALTinterwordspacing
B.~{Shi}, L.~{Zhu}, W.~{Cai}, N.~{Chen}, T.~{Shen}, P.~{Zhu}, F.~{Shu}, and
  J.~{Wang}, ``On {Performance Loss} of {DOA Measurement Using Massive MIMO
  Receiver} with {Mixed-ADCs},'' \emph{CoRR}, vol. abs/2104.02447, 2021.
  [Online]. Available: \url{https://arxiv.org/abs/2104.02447}
\BIBentrySTDinterwordspacing

\bibitem{2017Hybrid}
A.~Molisch, V.~Ratnam, S.~Han, Z.~Li, S.~Nguyen, L.~Li, and K.~Haneda,
  ``{Hybrid Beamforming} for {Massive MIMO: A Survey},'' \emph{IEEE Commun.
  Mag}, vol.~55, no.~9, pp. 134--141, 2017.

\bibitem{2013Nonlinear}
J.~Jensen, M.~Christensen, and S.~Jensen, ``{Nonlinear Least Squares Methods}
  for {Joint DOA} and {Pitch Estimation},'' \emph{IEEE Transactions on Audio,
  Speech, and Language Processing}, vol.~21, no.~5, pp. 923--933, 2013.

\bibitem{1990Maximum}
P.~Stoica and K.~Sharman, ``Maximum likelihood methods for direction-of-arrival
  estimation,'' \emph{IEEE Transactions on Acoustics, Speech, and Signal
  Processing}, vol.~38, no.~7, pp. 1132--1143, 1990.

\bibitem{2007An}
E.~Kelly, ``An {Adaptive Detection Algorithm},'' \emph{IEEE Trans. Aerosp.
  Electron. Syst}, vol. AES-22, no.~2, pp. 115--127, 1986.

\bibitem{2012Target}
R.~Niu, R.~Blum, P.~Varshney, and A.~Drozd, ``Target {Localization} and
  {Tracking} in {Noncoherent Multiple-Input Multiple-Output Radar Systems},''
  \emph{IEEE Trans. Aerosp. Electron. Syst}, vol.~48, no.~2, pp. 1466--1489,
  2012.

\bibitem{Goodman1963Statistical}
Goodman and R.~N., \emph{Statistical {Analysis Based} on a {Certain
  Multivariate Complex Gaussian Distribution (An Introduction)}}.\hskip 1em
  plus 0.5em minus 0.4em\relax Annals of Mathematical Statistics, 1963,
  vol.~34, no.~1.

\end{thebibliography}
\end{document}